\documentclass[twocolumn,showpacs,preprintnumbers,amsmath,amssymb]{revtex4}
\usepackage{graphicx}
\usepackage{dcolumn}
\usepackage{bm}

\begin{document}
\title{Entropic Penalties in Circular DNA Assembly}

\author{Marco Zoli}

\affiliation{School of Science and Technology - CNISM \\  University of Camerino, I-62032 Camerino, Italy \\ marco.zoli@unicam.it}

\date{\today}

\begin{abstract}
The thermodynamic properties of DNA circular molecules are investigated by a new path integral computational method which treats in the real space the fundamental forces stabilizing the molecule. The base pair and stacking contributions to the classical action are evaluated separately by simulating a broad ensemble of twisted conformations. We obtain, for two short sequences, a free energy landscape with multiple wells corresponding to the most convenient values of helical repeat. Our results point to a intrinsic flexibility of the circular structures in which the base pair fluctuations move the system from one well to the next thus causing the local unwinding of the helix.
The latter is more pronounced in the shorter sequence whose cyclization causes a higher bending stress. The entropic reductions associated to the formation of the ordered helicoidal structure are estimated.

\end{abstract}


\maketitle

\section*{I. Introduction}

DNA in cells is thermodynamically stabilized by electrostatic hydrogen bonds between bases on complementary strands and by strong van der Waals interactions between planar bases which stack on top of one another with rise distance $3.4$ \AA $\,$, along the double helical structure. As the phosphate groups in the molecule backbone are negatively charged, the strands tend to keep apart in order to minimize the electrostatic repulsion. Moreover, the polarity of the sugar-phosphate backbones render them hydrophillic hence, they are exposed to the aqueous solvent whereas the hydrophobic non-polar bases are buried in the core of the helix. When nucleotides are assembled by DNA polymerase into a regular structure, water molecules are expelled from the helix core and released in the surroundings. Accordingly, both the enthalpy and the entropy of the solvent increase thereby more than compensating the entropic penalties associated to the pairing and stacking of the bases. 
Hence, the order established in the growth of the open biophysical system is outbalanced by the disorder created in the environment consistently with the second law of thermodynamics \cite{boltz,avery}.  
While these effects which determine the free energy budget of the DNA-water system are understood, quantitative estimates of the entropic reductions ruling the helix formation \cite{schroe}  may offer important information regarding the interplay of pairing and stacking degrees of freedom in specific DNA sequences \cite{bloom,zocchi04}.

Circular DNA, the form generally found in prokaryotes and recently identified in mammalian cells \cite{dutta}, is currently investigated also for its technological potential, i.e. for the self-assembling of functional enzyme-DNA nanowires \cite{willner}. 
Cyclization of sequences with free ends has long been a major tool to infer information about torsional and bending properties of the helix \cite{shore}.  Ring closure probabilities in DNA with $N > 200$ base pairs (\textit{bps}) have been well described by worm-like chain (WLC) models  \cite{shimada} whereas substantial discrepancies between experiments and WLC predictions have been reported \cite{cloutier} for fragments with  $N \sim  100$ and less, which also show spontaneous loops formation. Such discrepancies have been resolved by recognizing that large angles can form between neighbor \textit{bps} along the molecule stack \cite{volo05}.  While large fluctuations determine the high looping rate observed in sequences whose length is less than the persistence length \cite{zocchi11,vafa,zocchi13,menon}, the intrinsic DNA flexibility could be ascribed either to transient kinks which unstack adjacent \textit{bps} or to the opening of small bubbles with separated strands \cite{yan04}.

Both mechanisms are incorporated in the statistical mechanical theory here proposed to evaluate the entropic pairing and stacking contributions to the formation of circular sequences. The idea underlying the computational method is the following: starting with an ensemble of $N$ free nucleotides, we \textit{first} constrain the centers of mass of the $N$  \textit{bps} to lie along a circle. As a\textit{ second} step, we consider the internal structure of the $i-th$ object which is made of two complementary bases tied by hydrogen bonds  \cite{demarco} with effective energies accounting for the phosphate electrostatic repulsion. The latter is screened by the counterions dissolved in the aqueous environment. \textit{Third}, we focus on adjacent objects along the molecule stack and model their interaction, formally described by a two-particles potential. Each of these steps introduces some degree of order (with respect to the original ensemble) which may depend both on the environment and on the precise conformation of the helicoidal molecule.
The calculation is based on a path integral method that substantially differs from the version presented in some recent studies \cite{io09,io11}. Precisely, the \textit{bps} fluctuations are thought of as time dependent paths whose interactions are now treated in the real space, not in the time lattice as previously assumed. This new mapping technique produces a more accurate mesoscopic description of the DNA structure and, in particular, of the stacking forces between adjacent nucleotides which are kept  at the experimental rise distance in the real space. Applying the theory to two short heterogeneous circles, we derive an overall picture which points to an intrinsic molecule flexibility sustained by sizeable base pair fluctuations with computed entropy values higher than those previously obtained.
Model and computational method are presented in Sections II and III, respectively. The thermodynamics results are discussed in Section IV, while some final remarks are reported in Section V.

\section*{II. Model}

We take an heterogeneous ensemble of $N$ purine-pyrimidine \textit{bps} with reduced mass $\mu$ and define $\textbf{r}_i$ as the center of mass coordinate for the $i-th$ base pair. In the finite temperature path integral formalism  \cite{io09,io11} the base pair displacements are treated as trajectories, $x_i(\tau)$,  depending on the imaginary time  $\tau=\,it$ and $t$ is the real time for the path evolution amplitude \cite{feyn}:

\begin{eqnarray}
\pm |\textbf{r}_i| \rightarrow  x_i(\tau) ; \, \, \, \, \tau \in [0 \,, \beta ] \,.
\label{eq:005}
\end{eqnarray}

$\beta=\,(k_B T )^{-1}$, $k_B$ is the Boltzmann constant and $T$ is the temperature.  It is remarked that the $i-th$ displacement in Eq.~(\ref{eq:005}) is not mapped onto a specific  $\tau_i$ whereas the mapping procedure in Refs. \cite{io09,io11} assumed, $|\textbf{r}_i| \rightarrow  x(\tau_i)$. As a consequence, $\tau$ is now kept as a free parameter and no partition of the $\beta$ length in $N$ intervals is carried out in the method presented in this paper.
The consequences will be emphasized in the following.

The base pair paths are expanded in Fourier series consistently with the closure condition,  $x_i(0)=\, x_i(\beta)$. Hence, we can define an integration measure $\oint {D}x$ over the space of the Fourier coefficients:

\begin{eqnarray}
& &x_i(\tau)=\, (x_0)_i + \sum_{m=1}^{\infty}\Bigl[(a_m)_i \cos(\omega_m \tau ) + (b_m)_i \sin(\omega_m \tau ) \Bigr] \, \nonumber
\\
& &\omega_m =\, \frac{2 m \pi}{\beta} \, \nonumber
\\
& &\oint {D}x \equiv {\frac{1}{\sqrt{2}\lambda_C}} \int d(x_0)_i \prod_{m=1}^{\infty}\Bigl( \frac{m \pi}{\lambda_C} \Bigr)^2 \int d(a_m)_i \int d(b_m)_i \, \, \nonumber
\\
\label{eq:01}
\end{eqnarray}

with $\lambda_C$ being the classical thermal wavelength. The $N$ \textit{bps} freely fluctuate in the space hence their partition function $Z_{k}$ is obtained by weighing
the contributions to the kinetic action $A_{k}[x]$ in the path phase space.
The measure $\oint {D}x$ normalizes the partition function $Z_{k}$  which decouples into a product of Gaussian integrals: 

\begin{eqnarray}
& & Z_{k}=\, \oint Dx \exp \bigl[- A_{k}[x] \bigr] =\, 1 \, \nonumber
\\
& &A_{k}[x]=\, \sum_{i=1}^{N} \int_{0}^{\beta} d\tau \biggl[ \frac{\mu}{2} \dot{x}_i^2(\tau) \biggr] \, .
\label{eq:001}
\end{eqnarray}

Eq.~(\ref{eq:001}) sets the zero for the free energy, $\beta^{-1} \ln Z_{k}$,  of the $N$ \textit{bps} ensemble. As a consequence of the real space mapping, the action is a dimensionless object.

\subsection*{A. Base Pairs on a Circle}

\begin{figure}
\includegraphics[height=7.5cm,width=9.5cm,angle=-90]{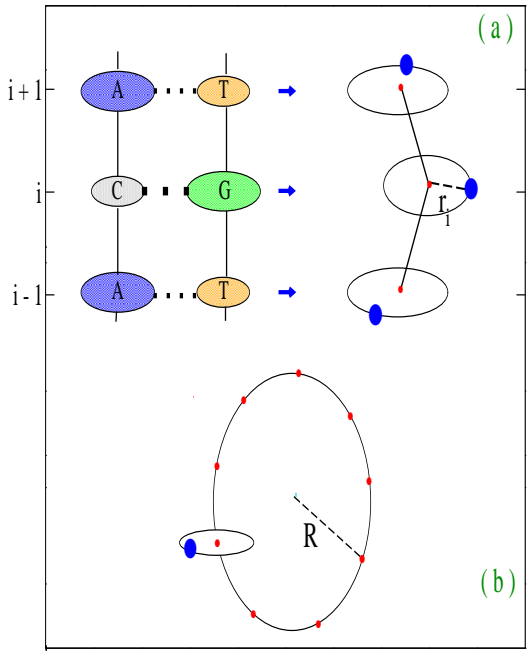}
\caption{\label{fig:1}(Color online)  
(a) Triplet of base pairs stacked along the DNA backbone. The blue filled circles are the base pair centers of mass.   $\textbf{r}_{i}$ is the fluctuational vector for the $i-th$ base pair. Adjacent centers of mass along the molecule axis are both twisted and bent. (b) $R$ is the radius of the DNA circle. For $|\textbf{r}_{i}|=\,0$, the fluctuational orbit shrinks into the red point which corresponds to the fluctuations free state for the $i-th$ base pair. }
\end{figure}

The centers of mass are now uniformly arranged on a circle with radius $R$, see Fig.~\ref{fig:1}, so that the rise distance between adjacent \textit{bps} is $2 \pi R / N$.
At this stage, no assumption is made about the inter-strand forces that bind the complementary bases and the intra-strand forces that bind the nucleotides along the stack. Every base pair may fluctuate with respect to the fundamental circle and describe an orbit spanned by $\textbf{r}_i$.   
Hence the general vector  $\textbf{t}_i$ for the for the $i-th$ center of mass is:

\begin{eqnarray}
& &\bigl({t }_i \bigr)_{x} =\, |\textbf{r}_i| \cos\phi_i \cos\theta_i \, \nonumber
\\
& &\bigl({t }_i\bigr)_{y} =\,(R + |\textbf{r}_i|\sin\theta_i) \cos\phi_i
\, \nonumber
\\
& &\bigl({t }_i\bigr)_{z} =\,(R + |\textbf{r}_i|) \sin\phi_i \,.
\,
\label{eq:004}
\end{eqnarray}

The polar angle, $\theta_i =\, (i - 1) 2\pi / h + \theta_S$, measures the $i-th$ base pair twisting around the molecule backbone and $h$ is the number of \textit{bps} per helix turn. By varying this number the computational method simulates a broad ensemble of possible twisted conformations.
The azimuthal angle, $\phi_i =\, (i-1){{2 \pi} / N} + \phi_S $,  represents the bending of the $i-th$ base pair. It is therefore assumed that the relative bending between adjacent \emph{bps} along the stack depends only on the circle length.
$\theta_S$ and $\phi_S$ are the twist and the bending of the first \emph{bp} in the sequence. As polynucleotide chains have a direction due to the chemistry of the bonds, the general model should account for a distribution of $\theta_S$ and $\phi_S$ values.
For a DNA circle, the fluctuational orbits defined by $i=\,1$ and $i=\,N+1$ overlap.

In analogy with Eq.~(\ref{eq:005}),  $\textbf{t}_i$ is mapped onto a time dependent path:

\begin{eqnarray}
& & |\textbf{t}_i| \rightarrow  \eta_i(\tau)\, , \,
\label{eq:005a}
\end{eqnarray}

with the path fluctuation $\eta_i(\tau)$ which, in the circular geometry, reads:

\begin{eqnarray}
& & \eta_i(\tau)=\, \bigl[R^2 + x_i(\tau)^2 + 2R |x_i(\tau)|f(\theta_i,\phi_i) \bigr]^{1/2} \, \nonumber
\\
& &f(\theta_i,\phi_i)=\,\sin\theta_i \cos^2 \phi_i + \sin^2 \phi_i  \,.
\label{eq:005b}
\end{eqnarray}

Accordingly, the partition function for the ensemble of $N$ centers of mass freely oscillating around the fluctuations free circle is given by:

\begin{eqnarray}
& & Z_{0}=\, \oint Dx \sum_{\theta_S, \phi_S} \exp \bigl[- A_{0}[\eta ] \bigr] \, \nonumber
\\
& &A_{0}[\eta ]=\, \sum_{i=1}^{N} \int_{0}^{\beta} d\tau \biggl[ \frac{\mu}{2} \dot{\eta }_i^2(\tau) \biggr] \, . \,
\label{eq:0001}
\end{eqnarray}

I emphasize that, in the path integral model, the action in Eq.~(\ref{eq:0001}) accounts for the bending energy cost due to the deformation which closes the array of $N$ 
\emph{bps} into a loop. It follows that, unlike $Z_{k}$, $Z_{0}$ does not decouple into a product of Gaussian integrals. For any finite $R$, from Eqs.~(\ref{eq:005a}),~(\ref{eq:0001}), one can compute the free energy associated to the cyclization of the $N$ \emph{bps} ensemble. With respect to this free energy level, the effects due to the hydrogen bond pairing and to the stacking can be evaluated.

Note that in the path integral method summarized by Eqs.~(\ref{eq:005}),~(\ref{eq:01}),~(\ref{eq:005a}), the rise distance along the molecule axis can be pinned to the experimental value for \textit{any} sequence length. This property could not be fulfilled by the previously used mapping technique which, operating a partition of the imaginary time axis, set a distance $\beta / N$ between neighboring \textit{bps} \cite{io09}. While this assumption had significantly reduced the computational time, it may have oversimplified the model and introduced some coarse approximations, mostly in the case of long sequences.

\subsection*{B. Inter-Strand Forces}

Next we look at the internal structure of the base pair and model the hydrogen bonds, two for AT-bps and three for GC-bps, by a Morse potential:  $\,V_{M}[ \eta _i(\tau)]=\, D_i \bigl[\exp(-b_i (\eta_i(\tau) - R)) - 1 \bigr]^2$. $D_i$ is the pair dissociation energy and the inverse length $b_i$ sets the potential range.  The hard core,  encountered by fluctuations smaller than $R$,  accounts for the repulsive electrostatic interaction between phosphate groups. The latter prevents the complementary bases from coming too close to each other. This physical criterion is implemented by discarding fluctuations such that, $\eta_i(\tau) - R < - \ln 2 / b_i$, which would yield a repulsive energy larger than $D_i$. Importantly, the ensemble of selected path fluctuations takes into account the type of base pair along the sequence.  On the other hand, for $\eta_i(\tau) \gg  R$, the pair mates would go infinitely apart without energy costs as there is no restoring force on the Morse plateau. However, once a pair breaks, the bases can bind to the surrounding solvent which accordingly stabilizes the strands. This effect can be modeled by a solvent potential: $V_{Sol}[\eta_i(\tau)]=\, - D_i f_s \bigl(\tanh((\eta_i(\tau) - R)/ l_s) - 1 \bigr)$ as proposed in Refs. \cite{collins,druk}. The factor $f_s$ is related to the counterions concentration in the solvent and $l_s$ tunes the width of the solvent barrier \cite{io11}. While both hydrogen bonds and solvent terms introduce some order in the \textit{bps} system, their contributions to the partition function are described by a one particle potential:

\begin{eqnarray}
& & Z_{1}=\, \oint Dx \sum_{\theta_S, \phi_S} \exp \bigl[- A_{1}[\eta ] \bigr] \, \nonumber
\\
& &A_{1}[\eta ]=\, \sum_{i=1}^{N} \int_{0}^{\beta} d\tau \biggl[ \frac{\mu}{2} \dot{\eta }_i^2(\tau) + V_{1}[ \eta _i(\tau)]  \biggr] \, \nonumber
\\
& &V_{1}[ \eta _i(\tau)]=\, V_{M}[ \eta _i(\tau)] + V_{Sol}[\eta_i(\tau)]\, . 
\label{eq:0002}
\end{eqnarray}

\subsection*{C. Intra-strand Stacking Forces}

A major computational task arises when stacking forces are introduced to pile up the nucleotides along the strands. While long range correlations should be taken into account in models for DNA coherent charge transport \cite{albu}, the thermodynamic quantities are generally dominated by short range correlations \cite{crothers0,crothers1}. This brings about a two particles potential that couples fluctuational paths between adjacent bases whose flat surfaces are perpendicular to the helix axis. $\pi - \pi$ electron interactions between bases are a primary source of duplex formation and stability  \cite{yakov,cooper}. Such forces are modeled by a potential (proposed in studies of DNA melting \cite{pey2}) here adapted to the path integral formalism, extended to the circular geometry and accounting for heterogeneity in the base sequence:

\begin{eqnarray}
& &V_{2}[ \eta _i(\tau), \eta _{i - 1}(\tau)]=\, K_S \cdot G_{i, i-1} \bigl(\eta_i(\tau) - \eta_{i-1}(\tau)  \bigr)^2 \, \nonumber
\\
& &G_{i, i-1}= \,1 + \rho_{i, i-1}\exp\bigl[-\alpha_{i, i-1}(\eta_i(\tau) + \eta_{i-1}(\tau)  - 2R)\bigr]  \, . \nonumber
\\ 
\label{eq:0003}
\end{eqnarray}

$K_S$ is the harmonic stiffness constant while $\alpha_{i, i-1}$ and $\rho_{i, i-1}$ weigh the effects of nonlinear path fluctuations. As the condition $\alpha_{i, i-1} < b_i$ is fulfilled, the range of the stacking is larger than that of the Morse potential. Accordingly, if $\eta_{i}(\tau)  - R \gg \alpha_{i, i-1}^{-1}$, the $i-th$ hydrogen bond is broken and the stacking coupling drops from \, $\sim K_S \cdot (1 + \rho_{i, i-1})$ to $\sim K_S$,  thus favoring the breaking of the adjacent base pair and the opening of local bubbles \cite{rapti,segal,metz11}. Then, by virtue of its nonlinear structure, the potential in Eq.~(\ref{eq:0003}) is adequate to account for those cooperative effects which may propagate along the molecule stack and cause even large fluctuational openings at high temperature \cite{io14a}.

Note that the backbone stiffness opposes the unstacking of the bases and maintains the strand stability. The finiteness of the stacking forces \cite{joy05} is ensured by the cutoffs in the integrations over the Fourier coefficients (in Eq.~(\ref{eq:01})) which truncate the path phase space. Adding the stacking to the classical action, the total partition function is eventually derived as:

\begin{eqnarray}
&&Z_{2}= \, \oint Dx\sum_{\theta_S, \phi_S} \exp \bigl[- A_{2}[\eta ] \bigr] \, \nonumber
\\
&&A_{2}[\eta ]= \, \sum_{i=1}^{N}  \int_{0}^{\beta} d\tau \biggl[ \frac{\mu}{2} \dot{\eta }_i^2(\tau) + V_{1}[ \eta _i(\tau)] + \, \nonumber
\\ 
& &V_{2}[ \eta _i(\tau), \eta _{i - 1}(\tau)] \biggr] \, .
\label{eq:0003a}
\end{eqnarray}

As a general consequence of the real space mapping technique, the classical action $A_{2}$ is obtained as a $\tau-$ integral of the kinetic \textit{plus} potential terms both depending on the base pair fluctuational paths.

\section*{III. Partition Function}

\begin{figure}
\includegraphics[height=7.5cm,width=9.5cm,angle=-90]{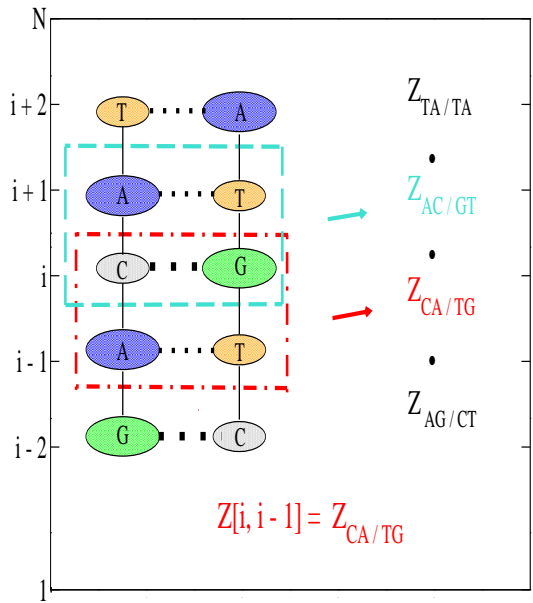}
\caption{\label{fig:2}(Color online) Diagram of the nearest neighbor method used in the computation of Eq.~(\ref{eq:04b}). The total partition function for $N$ nucleotides is obtained as a product of the pairing and stacking contributions stemming from a quadruplet of nucleotides. Every base pair, e.g. the $i-th$ CG in the diagram, participates to two neighbor blocks along the molecule stack. }
\end{figure}

For computational purposes,  Eq.~(\ref{eq:0003a}) can be rewritten as a product of factors involving two pairs of adjacent bases as schematically shown in Fig.~\ref{fig:2}. Each base is bound to a sugar-phosphate group (not drawn) whose effect is considered through the effective model parameters. 
Hence $Z_2$ reads:

\begin{eqnarray}
& &Z_2=\,\biggl( \prod_{i=1}^{N}  Z[i,\,i - 1] \biggr)^{1/2} \, \nonumber
\\
& &Z[i,\,i - 1]=\, \sum_{\theta_S, \phi_S} \oint  \textsl{D}(\bar{x})_i  I[i] \cdot \oint  \textsl{D}(\bar{x})_{i - 1} J[i,\,i - 1] \,\, \nonumber
\\
& &I[i]=\, \exp\biggl[- \int_{0}^{\beta} d\tau \biggl( \frac{\mu}{2} \dot{\eta }_i^2(\tau) + V_{1}[ \eta _i(\tau)] \biggr) \biggr] \, \nonumber
\\
& &J[i,\,i - 1]=\, \exp\biggl[- \int_{0}^{\beta} d\tau V_{2}[ \eta _i(\tau), \eta _{i - 1}(\tau)] \biggr] \, . \,
\label{eq:04b}
\end{eqnarray}

$\oint \textsl{D}(\bar{x})_i$ indicates multiple integrals over normalized Fourier coefficients for the $i-th$ base pair. The square root in the first of Eqs.~(\ref{eq:04b}) is due to the fact that each base pair enters the calculation of two adjacent blocks. This problem is formally analogous to that encountered in the statistical mechanics of Ginzburg-Landau fields whose partition function has been evaluated by transfer integral techniques \cite{scalap}. A similar approach has been applied to a mesoscopic DNA Hamiltonian to derive the melting profiles in terms of the eigenvalues and eigenstates of a transfer integral equation  \cite{zhang}. Unlike transfer integral methods, the path integral computation directly sums over those sets of Fourier coefficients in Eq.~(\ref{eq:01}) which define base pair paths consistent with the physical requirements of the model potential as described in Section II. A detailed description of the computational method is found in Refs.\cite{io11}.

Switching off $V_{2}$ in Eqs.~(\ref{eq:04b}), one recovers $Z_1$ of Eq.~(\ref{eq:0002}). Switching off both $V_{2}$ and $V_{1}$, one gets $Z_0$ of Eq.~(\ref{eq:0001}).
Accordingly, the pairing and stacking contributions to the partition function, for a specific sequence, are evaluated separately. This is done by varying the helical repeat ($h$) in a broad range of values which correspond to a set of possible helicoidal conformations. 

As any sampled conformation may in principle correspond to a state of thermal equilibrium, the path integration in Eq.~(\ref{eq:04b}) is performed for all twisted conformations in the set. For any $h$, the program simulates an ensemble of molecular states and every state is given, in turn, by a configuration of $N$ \textit{bps} fluctuations selected in the path phase space. The calculation includes in the partition function about $4.5 \cdot 10^7$ paths for each base pair and for any choice of twisted conformation.
This marks a substantial methodological difference with respect to the stochastic nature of Monte Carlo simulations \cite{ares} which assume an initial distribution of base pair states and search for thermal equilibrium by inducing random changes in the microscopic states of successive trial conformations \cite{note0}.

While specific experimental information regarding the value of $h$ in short circular sequences are not available, the analysis here proposed intends to have a predictive value.
It is remarked that $h$ measures the average torsional stress of the molecule whereas it does not precisely locate the occurrence of helical disruptions along the sequence which, instead, can be experimentally detected by techniques such as fluorescence correlation spectroscopy \cite{bonnet} and theoretically predicted by analysis of bubble profiles \cite{benham,hwa,palmeri,metz10,singh}.

The theory is applied to two circular sequences with almost equal content of GC-\textit{bps} and AT-\textit{bps} but different sizes, $N=\,66 \, , \,86$, which have been prepared as detailed in ref.\cite{volo08}. These circles have been recently studied to test the effects of the bending and torsional stress on the stability of the helix \cite{io14a}. Their thermodynamical properties are now examined at the light of the new computational approach.
The model parameters are taken as: $D_{AT}=\,30\, meV$,  $D_{GC}=\,45\, meV$,   $b_{AT}=\,1.7$ \AA$^{-1}$,  $b_{GC}=\,2$ \AA$^{-1}$, $f_s=\,0.1$, $l_s=\,0.5$ \AA, $K_S =\,10 \,meV$ \AA$^{-2}$, $\rho_{i, i-1} \in [1,4]$,  $\alpha_{i, i-1} \in [0.2, 0.4]$ \AA$^{-1}$, consistently with Ref.\cite{io14a}. Somewhat smaller values for $b_{i}$ are here chosen in line with Ref.\cite{zdrav} and with general studies of mesoscopic models fitting mechanical opening experiments on single molecules  \cite{heslot}. The elastic force constant $K_S$ is taken in the low range of the reported values \cite{fenn,eijck} as it seems appropriate to capture the intrinsic flexibility displayed by the helix  at the microscopic level \cite{weber13,mazur}.

A simulation, e.g. for the $N=\,86$ sequence at room temperature, takes about 190 hours on a workstation (Intel Xeon E5-1620 v2, 3.7GHz processor), a remarkably longer time than that required by our previous method.

\section*{IV. Free Energy and Entropy}

Arranging the $N$ centers of mass into a loop, we obtain free energy and entropy for the bent configuration from Eqs.~(\ref{eq:0001}) as: $F_{0}=\, - \beta^{-1} \ln Z_{0} $ and $S_{0}=\, - \partial F_{0} /\partial T \bigl|_V $. With respect to these quantities, we evaluate the effects of the pairing and stacking potentials.

Free energy and entropy for the circular sequence of $N$ base pairs, with pairing interactions, are calculated via Eqs.~(\ref{eq:0002}) as: $F_{1}=\, - \beta^{-1} \ln Z_{1} $ and $S_{1}=\, - \partial F_{1} /\partial T \bigl|_V $, respectively.
Total free energy and entropy for the system, with {both} pairing and stacking forces, are obtained from Eqs.~(\ref{eq:04b}) by computing: $F_{2}=\, - \beta^{-1} \ln Z_{2} $ and $S_{2}=\, - \partial F_{2} /\partial T \bigl|_V $, respectively.

\begin{figure}
\includegraphics[height=7.5cm,width=9.5cm,angle=-90]{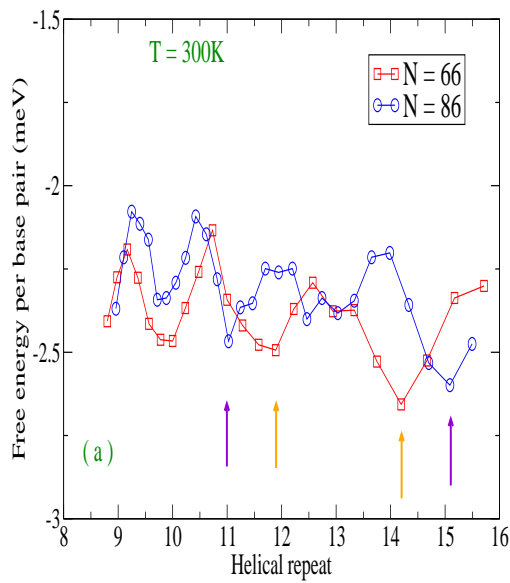}
\includegraphics[height=7.5cm,width=9.5cm,angle=-90]{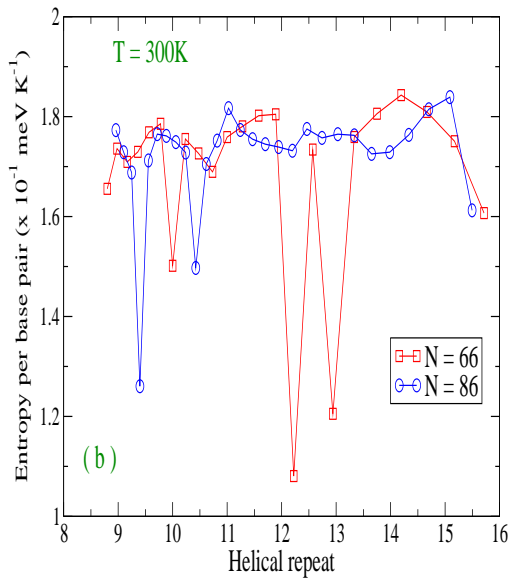}
\caption{\label{fig:3}(Color online)  (a) Total free energy per base pair, $F_2$ in the text;  (b) total entropy per base pair, $S_2$ in the text. Both are calculated, at room temperature, for a set of values of helical repeat. Two circular sequences, with $N=\,66, \,86$, are compared. }
\end{figure}

$F_{2}$ and  $S_{2}$ are plotted in Figs.~\ref{fig:3} for the two heterogeneous sequences. The calculation is performed at room temperature for an ensemble of helicoidal conformations by varying $h$ in a broad range. The physically significant window of $h$ values is presented in the plots. 

While repeats of $\sim 10.4 - 11.3$ \textit{bps} per turn are typically found in distributions of closed molecules \cite{depew,cozza},  changes in the environmental conditions (e.g., salt concentration) and thermal fluctuations may affect the supercoiling degree and the helix twist \cite{benham93b,marko95}.

The free energy landscape (Fig.~\ref{fig:3}(a))  presents multiple wells signaling that, while some conformations are energetically favored, base pair fluctuations can drive the system out of a well causing the molecule to adopt a new state. 
Fig.~\ref{fig:3}(b) shows the entropic profile with the high entropy states corresponding to the free energy wells.
The two deepest free energy minima for each sequence are pointed out by arrows in Fig.~\ref{fig:3}(a). Both circles take helical repeats in the usual range $h \in [\sim 10, 12]$ but the lowest conformational free energies are found at larger values (i.e. $14.2$ and $15.1$). This indicates that both sequences experience significant \textit{bps} disruptions which untwist the helix \cite{lavery}. 
For instance, a shift in the helical repeat from $h=\,12$ to $h=\,14.2$ in the $66-$ \textit{bps} sequence amounts to reduce the average twist from $30^o$ 
to  $\sim 25^o$.

Some significant differences are found between the profiles of the two circles. Essentially, the $66-$ \textit{bps} sequence presents: \textit{ i)} a free energy minimum at larger $h$ in the standard range, \textit{ii)} a deeper (and therefore more favorable) free energy minimum  for $h$ in the untwisted range. This is physically understood by observing that,
in the untwisted conformation, the breaking of base pair bonds is more likely to occur in shorter sequences as  the bending stress due to cyclization is higher. 

These findings are consistent with the experiments \cite{volo08} and, in this regard, also with the results obtained in Ref.\cite{io14a}. 
While the latter work found, at room temperature, the free energy minima in the range $h \in [\sim 10, 11]$, it did not however capture the minima for large $h$ which  instead appear  in Fig.~\ref{fig:3}(a). 

The multiple wells free energy profile is thus peculiar of the more advanced computational technique presented in this paper. Also the values of entropy per base pair
are larger than those obtained by the previous method \cite{io11,note}, although they remain smaller by a factor five than the experimental estimates \cite{santa,blake,russu1}.
Nevertheless, this discrepancy can be at least partly reconciled by further tuning the model parameters and, specifically, by softening the hydrogen bond potential which, in turn, yields higher entropy values.

\begin{figure}
\includegraphics[height=7.5cm,width=9.5cm,angle=-90]{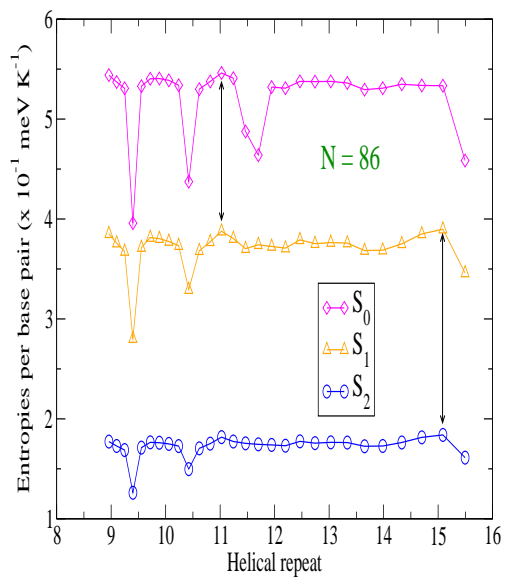}
\caption{\label{fig:4}(Color online)   
Entropic contributions, $S_0, \, S_1, \, S_2$ in the text, calculated at room temperature for the $N=\,86$ sequence. The arrows mark the entropic reductions, $S_0 - S_1$ and $S_1 - S_2$, at the values of helical repeat which maximize $S_0$ and $S_1$. }
\end{figure}

Fig.~\ref{fig:4} plots, for the circle with  $N=\,86$, the terms  $S_0$, \, $S_1$ and $S_2$ as a function of $h$ to give a quantitative estimate of the entropic reductions due to the one particle and two particles potential, respectively.

First,  the centers of mass are arranged along a circle. This has the effect to favor the conformation with $h \sim 11 \,$ for which $S_0$ gets the maximum. 
Next, we consider the hydrogen bond interactions between the pair mates and strands recombination with the solvent. This leads to two interesting effects: \textit{i)}  $S_1$ is reduced with respect to $S_0$, \textit{ii) }  the maximum for $S_1$ is shifted at larger $h$ thus suggesting that hydrogen bonds are a major source of flexibility in the molecule. 
Finally, we add the stacking potential and get a further, sizeable reduction for $S_2$ which makes $S_1 - S_2$ even larger than $S_0 - S_1$. 
Piling up the nucleotides along the bent molecule stack has therefore the effect to strongly order and stabilize the circular structure which albeit can be found in a set of possible conformations with almost equivalent free energies \textit{and} different helical pitches.  
An analogous behavior is obtained for the $N=\,66$ sequence.

\section*{V. Conclusions }

This paper presents a path integral computational method for the thermodynamics of circular DNA  which can be applied to sequences of any length, at any temperature.
The base pair thermal fluctuations are described by paths whose radial and angular degrees of freedom are included in the partition function. 
The latter is directly computed in the path phase space by summing over those base pair fluctuations  which are consistent with the mesoscopic potential modeling the essential interactions in the bent helicoidal conformations.

The method constitutes a significant advancement with respect to the path integral analysis  \cite{io09} developed in the last years.

The main property of the present technique lies in the fact that the stacking forces, between adjacent base pairs along the molecule backbone, are treated in the real space thus respecting the actual rise distance of the duplex. This feature was not shared by the previous theoretical scheme which performed a mapping of the base pairs onto the imaginary time lattice therefore modeling the stacking forces as a time-retarded interaction. About $4.5 \cdot 10^7$ paths for each base pair have been included in the partition function to simulate the fluctuational states along the sequence.

The fluctuational paths depend both on the radial distance between the base pair mates and on the polar angle peculiar of the helicoidal conformation whereas the azimuthal angle is set by the number of base pairs arranged along the circle.
A broad spectrum of twisted conformation has been sampled to obtain the free energy profiles at room temperature.
Certainly, the improvement achieved in the path integral technique comes at the price of a much longer computational time.

Applying the theory to the evaluation of the thermodynamic properties of two circular sequences, it is shown that the improved computational method yields a novel free energy landscape with a set of minima corresponding to specific values of the helical repeat.  As the base pair fluctuations can drive the molecule from one free energy minimum to the next, the theoretical picture suggested by our results points to a high degree of flexibility for the circular sequences which have the capability to adopt different helicoidal conformations. 
Untwisted molecules, with larger helical repeat than the regular structure, are likely to exist even at room temperature although the overall helix stability is maintained. 

We find that the helix untwisting is higher in shorter sequences as detected experimentally and therefore the disruptions of some hydrogen bonds appear as a strategy to release the stress associated to the curvature of the molecule axis.  Our results indicate that the model potential is sufficiently structured to account for the bending energy associated to the stiffness of short sequences which nonetheless maintain a substantial flexibility at the level of the base pairs.

Analyzing the entropy as a function of the helical repeat, we could quantitatively discern the contributions stemming from the base pairing and from the coupling between bases along the molecule stack. Both the former and mostly the latter, stabilize the helix and sensibly reduce the entropy values with respect to the loop configuration in which the base pairs do not experience any internal force. The total entropies per base pair turn out to be larger than those obtained by the previous path integral method but remain somewhat lower than the experimental values. These residual discrepancies may be at least partly ascribable to the fact that a constant bending angle between adjacent nucleotides has been assumed in our mesoscopic model whereas a more general theoretical framework should drop also this constraint thus allowing for a larger local flexibility along the stack. It should be however remarked that, for short molecules as those here considered, the stacking axis lies essentially in a plane and the linking number of the circular form coincides with the twist. Accordingly large fluctuations in the base pair azimuthal angles are unlikely to occur mostly at room temperature. While the body of the results here presented 
should hold also in the presence of azimuthal base pair fluctuations,
further work is underway to achieve an even more refined computation of the thermodynamical properties of circular DNA and next, of linear sequences.


\begin{thebibliography}{widest-label}
\bibitem{note}
Given the same sequence and same parameters, the room temperature entropy computed by the present method is larger by a factor three than that previously found.
\end{thebibliography}
\end{document}